\documentstyle[aps,prl,twocolumn]{revtex}

\input{tcilatex}

\begin{document}

%\preprint{Draft}
{\bf Comment on ``Magnetoresistance Anomalies in Antiferromagnetic YBa}$_{2}$%
{\bf Cu}$_{3}${\bf O}$_{6+x}${\bf : Fingerprints of Charged Stripes''}

In a recent Letter\cite{ando}, Ando {\it et al}. discovered an anomalous
magnetoresistance(MR) in hole doped antiferromagnetic YBa$_{2}$Cu$_{3}$O$%
_{6+x}$, which they attributed to charged stripes, i.e., to segregation of
holes into lines. \ In this Comment we show that the experiments, albeit
being interesting, do not prove the existence of stripes. \ In our view the
anomalous behavior is due to an ({\it a,b}) plane anisotropy of the
resistivity in the bulk and to a magnetic field dependent antiferromagnetic
(AF) domain structure. It is unlikely that domain walls are charged stripes.

The main experimental findings\cite{ando} were the following: the anomalous
MR i.) appears only with magnetic field in the ({\it a,b}) plane, ii.)
saturates at magnetic fields of a few Tesla, iii.) changes sign when the
magnetic field turns from parallel to perpendicular to the current, iv.)
disappears gradually as the temperature is raised to the Ne\'{e}l
temperature($T_{N}$), v.) depends on magnetic field history below 20 K. To
explain the observations it was proposed that holes order into an array of
stripes which act as current paths. The array of holes was supposed to be
charged, ferromagnetic and to reorder in small magnetic fields.

The AF domain structure has been recently studied\cite{janossy99} in detail
in insulating YBa$_{2}$Cu$_{3}$O$_{6+x}$ single crystals doped with 1\% Gd.
Substitution of Gd serves as an ESR probe, and physical properties are
unaffected. Differently oriented AF domains appear as distinct series of
lines in the ESR spectra of Gd$^{3+}$ ions. In the study of Ando et al.\cite
{ando} crystals had an oxygen concentration of $x\simeq 0.3$ while in the
ESR study $x<0.15$. We assume, however, that the domain structures are
similar in the two cases. According to the ESR study, the easy axis of the
AF order is along [100] and in zero magnetic field high quality crystals
consist of equal amounts of AF domains oriented along the two possible easy
axes. [110] is a hard axis in the (a,b) plane. Magnetic fields in the ({\it %
a,b}) plane reorient domains. At $T=20$ K a field of 5 T applied along an (%
{\it a,b}) plane principle axis is enough to turn practically all domains
perpendicular to the field. The component of the magnetic field along {\it c}
does not affect the domain structure. The anisotropy of the magnetic
susceptibility of the domains is the driving force of the reorientation. The
ESR showed a temperature independent domain structure between 10 and 150 K.

Above $T_{N}$ the crystal structure of lightly hole doped YBa$_{2}$Cu$_{3}$O$%
_{6+x}$ is tetragonal. In the AF ordered state it has a small orthorhombic
distortion along [100] due to coupling of the crystal lattice with the
ordered Cu(2) magnetic moments. We suggest that the lowering of the symmetry
leads to an anisotropy of the (bulk) resistivity in the ({\it a,b}) plane.
The resistivity is larger when current is parallel to the sublattice
magnetization and smaller when it is perpendicular. In the absence of a
magnetic field, the average resistivity is measured. Large magnetic fields
wipe out unfavored domains and the anisotropy appears in the resistance. The
anisotropy disappears gradually as $T_{N}$ is approached. The symmetry of
the in-plane MR reflects the symmetry of the orthorhombic distortion of the
lattice.

Thus there is no need for ferromagnetically ordered charged arrays of holes
to explain the MR. The question of the nature of the domain walls remains,
however, unanswered. Neither ESR nor MR experiments distinguish between AF
domain walls parallel or perpendicular to the c axis. As explained in Ref. 
\cite{janossy99}, neutral domain walls perpendicular to {\it c} are more
likely to be the cause of the observed magnetic field dependence of the
domain structure.

The experiments do not rule out that walls consist of charged stripes of
holes in the ({\it a,b}) plane but are no confirmation either. It is
difficult to see why would relatively small magnetic fields change the array
of charged lines since Coulomb repulsion would rend such an arrangement
extremely rigid.

Support by the Hungarian State Grants OTKA-T029150 and HAS-AKP is
acknowledged.

\bigskip

Andr\'{a}s J\'{a}nossy$^{\ddag }$, Ferenc Simon, and Titusz Feh\'{e}r

Institute of Physics, Technical University of Budapest, P.O. BOX 91, H-1521
Budapest, Hungary



\begin{references}
\bibitem[\ddag ]{email}  Corresponding author: atj@power.szfki.kfki.hu.

\bibitem{ando}  Y. Ando, A. N. Lavrov, and K. Segawa, Phys. Rev. Lett. {\bf %
83}, 2813 (1999).

\bibitem{janossy99}  A.\ J\'{a}nossy {\it et al.}, Phys. Rev. B{\bf \ 59},
1176 (1999).
\end{references}
\end{document}